\documentclass[%
 aip,
 amsmath,amssymb,
 reprint,%
]{revtex4-1}

\usepackage{graphicx}
\usepackage{dcolumn}
\usepackage{bm}
\usepackage{hyperref}
\usepackage[utf8]{inputenc}
\usepackage[T1]{fontenc}
\usepackage{mathptmx}
\usepackage{etoolbox}

\usepackage{xcolor}

\makeatletter
\def\@email#1#2{%
 \endgroup
 \patchcmd{\titleblock@produce}
  {\frontmatter@RRAPformat}
  {\frontmatter@RRAPformat{\produce@RRAP{*#1\href{mailto:#2}{#2}}}\frontmatter@RRAPformat}
  {}{}
}%
\makeatother
\begin{document}


\title{Systematic investigation of the generation of luminescent emitters in hBN via irradiation
engineering}

\author{Pooja C. Sindhuraj}
\affiliation{Department of Electrical and Photonics Engineering, Technical University of Denmark, 2800
Kgs. Lyngby, Denmark}
\author{José M. Caridad}
\affiliation{Departmento de Física Aplicada, Universidad de Salamanca, Salamanca 37008, Spain}
\affiliation{Unidad de Excelencia en Luz y Materia Estructurada (LUMES), Universidad de Salamanca,
Salamanca 37008, Spain}
\author{Corné Koks }
\affiliation{Department of Electrical and Photonics Engineering, Technical University of Denmark, 2800
Kgs. Lyngby, Denmark}
\affiliation{NanoPhoton – Center for Nanophotonics, Technical University of Denmark, 2800 Kgs.
Lyngby, Denmark}
\author{Moritz Fischer}
\affiliation{Department of Electrical and Photonics Engineering, Technical University of Denmark, 2800
Kgs. Lyngby, Denmark}

\author{Denys I. Miakota}
\affiliation{Department of Electrical and Photonics Engineering, Technical University of Denmark, 2800
Kgs. Lyngby, Denmark}
\author{ Juan A. Delgado-Notario}
\affiliation{Departmento de Física Aplicada, Universidad de Salamanca, Salamanca 37008, Spain}
\affiliation{Unidad de Excelencia en Luz y Materia Estructurada (LUMES), Universidad de Salamanca,
Salamanca 37008, Spain}
\author{Kenji Watanabe}
\affiliation{Research Center for Electronic and Optical Materials, National Institute for Materials Science, 1-1 Namiki, Tsukuba 305-0044, Japan}
\author{Takashi Taniguchi}
\affiliation{Research Center for Materials Nanoarchitectonics, National Institute for Materials Science, 1-1 Namiki, Tsukuba 305-0044, Japan}
\author{Stela Canulescu}
\affiliation{Department of Electrical and Photonics Engineering, Technical University of Denmark, 2800 Kgs. Lyngby, Denmark}
\author{Sanshui Xiao}
\affiliation{Department of Electrical and Photonics Engineering, Technical University of Denmark, 2800
Kgs. Lyngby, Denmark}
\affiliation{NanoPhoton – Center for Nanophotonics, Technical University of Denmark, 2800 Kgs.
Lyngby, Denmark}
\author{Martijn Wubs}
\affiliation{Department of Electrical and Photonics Engineering, Technical University of Denmark, 2800
Kgs. Lyngby, Denmark}
\affiliation{NanoPhoton – Center for Nanophotonics, Technical University of Denmark, 2800 Kgs.
Lyngby, Denmark}
\author{Nicolas Stenger}
\altaffiliation[Corresponding author. Email: ]{niste@dtu.dk}
\affiliation{Department of Electrical and Photonics Engineering, Technical University of Denmark, 2800
Kgs. Lyngby, Denmark}
\affiliation{NanoPhoton – Center for Nanophotonics, Technical University of Denmark, 2800 Kgs.
Lyngby, Denmark}

\date{\today}

\begin{abstract}
Hexagonal boron nitride (hBN), a two-dimensional (2D) material, garners interest for hosting bright quantum emitters at room temperature. 
A great variety of fabrication processes have been proposed with various yields of quantum emitters. In this work, we study the influence of several parameters, such as irradiation energy, annealing environment, and the type of hBN, on the emitter density in hBN. Our results show (i) high emitter density with oxygen irradiation at 204 eV, (ii) post-annealing in carbon-rich atmospheres significantly increases emitter density, reinforcing carbon's potential role, (iii) no significant effect of oxygen pre-annealing, and (iv) a slightly increased emitter density from hBN crystals with lower structural quality. Although the precise origin of the emitters remains unclear, our study shows that oxygen irradiation and subsequent inert annealing in a carbon-rich environment play a crucial role in emitter generation, while the other processing parameters have a smaller influence. As such, our systematic study and findings show relevant advances towards the reproducible formation of visible-frequency quantum emitters in hBN. 
\end{abstract}

\maketitle

\section{Introduction}\label{sec:Introduction}
\begin{figure*}
    \centering
    \includegraphics[width=0.9\linewidth]{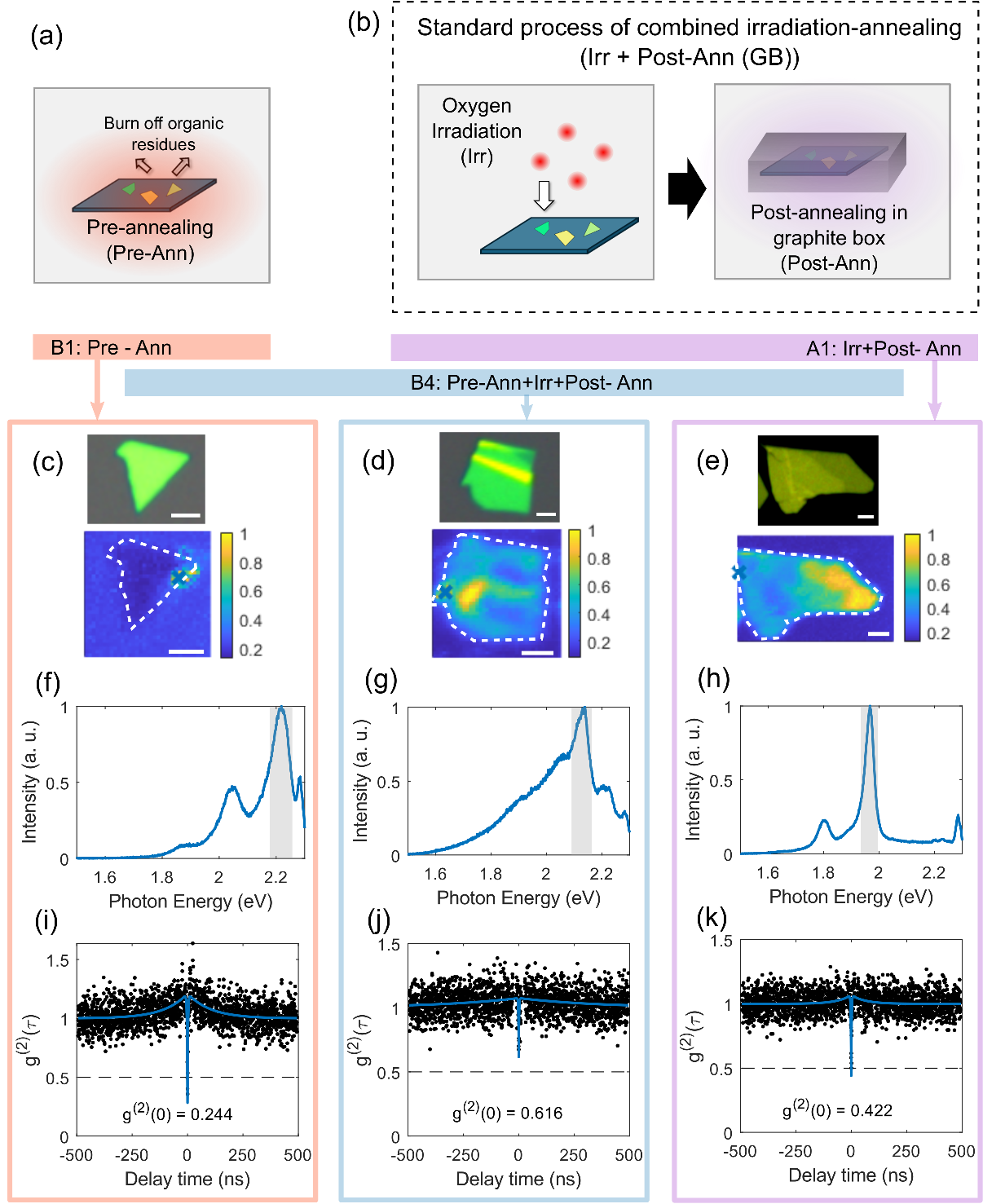}
    \caption{Schematic representation of the (a) pre-annealing treatment of exfoliated flakes in oxygen/argon atmosphere done to remove any organic residues, (b) standard process involving the oxygen irradiation of exfoliated flakes followed by post-annealing in nitrogen in a graphite box. Optical image (top) and photoluminescence map (bottom) of flakes undergoing (c) pre-annealing alone (B1: Pre-Ann), (d) pre-annealing followed by the standard process (B4: Pre-Ann+Irr+Post-Ann), (e) standard process (A1: Irr+Post-Ann). The white scale bar inside represents 5$\mu$m, and the color bar represents the normalized integrated PL counts. (f-h) Spectrum of a representative emitter generated in the corresponding processes. The location of the emitter is denoted as a blue cross in the PL maps. (i-k) g\textsuperscript{(2)}($\tau$) function of the emitters shown in (f), (g), and (h), respectively, from the spectral region shaded in grey. Treatment combinations undergone by different sample batches and the corresponding emitter densities are also listed in Table~\ref{tab:table1}.}
    \label{fig:fig1}
\end{figure*}

\begin{table*}
\caption{\label{tab:table1} The table gives the details of different batches, treatments performed for each batch, and their recorded emitter density. All the irradiation processes mentioned in the table are done at $\sim$204 eV. The following short names are used for different processes, and the specific parameters for the indicated process for a given batch are explained in detail under the Methods or Results section. Batches that have undergone the standard process are highlighted in bold.\\
   Irr: Irradiated, Pre-Ann: Pre-annealing, Post-Ann: Post-annealing, GB: Graphite box, QB: Quartz boat, Type I: HQGraphene hBN, Type II: HPHT NIMS hBN. }
\begin{ruledtabular}
\begin{tabular}{cccc}
& &\multicolumn{2}{c}{Emitter Density ($\#/\mu {m}^2$)}\\
Batch&Treatment&Type I&Type II\\ \hline
\\
\textbf{A1}&\textbf{Irr + Post-Ann(GB); Standard process}& \textbf{0.0192$\pm$0.0019} & -\\
      A2&Irr + Post-Ann(QB)& 0.0106$\pm$0.0022 & -\\
      A3&Post-Ann(GB)& 0.0041$\pm$0.0015 & -\\
     \hline \\
     B1&Pre-Ann& 0.0093$\pm$0.0022 & 0.0045$\pm$0.0010\\
     B2&Pre-Ann + Irr& 0.0085$\pm$0.0019 & 0.0046$\pm$0.0013\\
     B3&Pre-Ann + Post-Ann(GB)& 0.0067$\pm$0.0013 & 0.0118$\pm$0.0012\\
     \textbf{B4}&\textbf{Pre-Ann + Irr + Post-Ann(GB)}& \textbf{0.0167$\pm$0.0019} & \textbf{0.0138$\pm$0.0016} \\
\end{tabular}
\end{ruledtabular}
\end{table*}

Quantum emitters (QEs) in hexagonal boron nitride (hBN) have attracted significant interest since their discovery in 2016 \cite{Tran2016QuantumMonolayers}. 
These QEs exhibit bright and stable emission \cite{}, even at room temperature, making them promising for quantum communication, sensing, and metrology applications without the need for costly and power-intensive cryogenic environments \cite{OBrien2009PhotonicTechnologies}. 
They emit light across a broad range of the electromagnetic spectrum, from near-infrared (NIR) to ultraviolet (UV) frequencies, and some even show magnetic properties at room temperature, which is particularly useful for sensing applications \cite{Tran2017RobustNitride, Jungwirth2016TemperatureNitride, Gottscholl2020InitializationTemperature, Stern2022Room-temperatureNitride, Stern2024AConditions}. 

A great number of generation mechanisms have been proposed for these quantum emitters including edge creation\cite{Ziegler2019DeterministicCreation}, focused ion beam irradiation\cite{Glushkov2022EngineeringWater, Liu2023SingleBeam, Klaiss2022UncoveringFabrication, Sarkar2023IdentifyingIrradiation}, laser irradiation\cite{Hou2018LocalizedNitride, Gan2022Large-ScaleNitride}, localized strain\cite{Proscia2018Near-deterministicNitride}, bottom-up growth on patterned substrates\cite{Li2021ScalableNitride}, and electron beam irradiation\cite{Fournier2021Position-controlledNitride, Gale2022Site-SpecificNitride, Kumar2023LocalizedNitride}. 
All these methods suggest the formation of crystallographic defects in the hBN lattice, analogous to vacancies in diamond \cite{Kurtsiefer2000StablePhotonsb} and other bulk semiconductors\cite{Morfa2012Single-photonDefects, Soltamov2012Room6H-SiC}. This leads to the attribution of these emitters to the presence of defects or impurities in the hBN lattice\cite{Tran2016QuantumMonolayers, Qiu2021AtomicFunctionalitiesd, Koperski2020MidgapNitride, MacKoit-Sinkeviciene2019CarbonNitride, Xu2023GreatlyNanocavity}. Recent studies, however, have also suggested that polycyclic aromatic hydrocarbons (PAHs)\cite{Toninelli2021SingleTechnologies} may create bright emitters at the interface between hBN flakes and their supporting substrate when subjecting the samples to an inert annealing step\cite{Neumann2023OrganicMica}. In line with this observation, other recent works have shown that hBN serves as an excellent substrate for molecular QEs; for example, terrylene molecules adsorbed on hBN flakes showed narrower linewidths and better spectral stability than on other host materials \cite{Smit2023SharpSurface, Han2021PhotostableTemperatureb,deHaas2024ChargeNitride}. Importantly, many of the aforementioned methods for QEs share common processing steps and, moreover, the presence of natural defects in hBN crystals or residual organic molecules cannot be fully excluded in any of the fabrication approaches\cite{Ziegler2019DeterministicCreation, Glushkov2022EngineeringWater, Hou2018LocalizedNitride, Proscia2018Near-deterministicNitride,Kumar2023LocalizedNitride, Neumann2023OrganicMica}.
A clear understanding of the origin of the emitters is therefore lacking in the field, a fact which is further accentuated by the absence of detailed studies reporting the impact of the different steps of parameters during fabrication.

In this work, we undertake a systematic investigation of the generation mechanism of emitters with zero phonon line (ZPL) emission energies between 1.6 and 2.3 eV through irradiation with low-energy, high-fluence oxygen atoms, followed by post-annealing steps\cite{Fischer2021ControlledEngineeringb, Fischer2023CombiningOrigin, Vogl2019AtomicNitride}. In particular, we exhaustively investigate the role of commonly used parameters such as (i) irradiation energy, (ii) post-annealing with a carbon-rich environment, (iii) pre-annealing with an oxygen-rich environment, and (iv) the type of hBN crystals that are used. Whereas the precise origin of the emitters that are observed is beyond the scope of this study, our observations point towards certain factors, such as the influence of the actual irradiation step. Comparing the various parameters by their emitter yields allows us to gain control over the emitter generation. This is essential information for designing QE-based devices for quantum information processing, sensing, and metrology in a reproducible manner.

\section{Methods}\label{sec:Methods}

\subsection{Sample Preparation}
We exfoliate hBN flakes from bulk crystals purchased from HQGraphene
onto cleaned SiO\textsubscript{2}/Si substrates. 
The substrate chips are cleaned with 2 min sonication in acetone and 1 minute sonication in IPA prior to the exfoliation. 
Exfoliated samples are irradiated with oxygen within a Reactive Ion Etch Chamber (RIE). 
The RIE generates highly directional oxygen plasma targeted to the sample at a flow of 40 sccm, maintaining a 25 mTorr pressure inside the chamber. 
The plasma power that can be manually set determines the kinetic energy at which the plasma reaches the sample. 
For the plasma powers at 20 W, 40 W, and 60 W, kinetic energies at 120 eV, 204 eV, and 280 eV are achieved, respectively, with an error of $\pm$5 eV. 
The irradiation is done for 10 seconds in all cases and at room temperature. We chose 10 seconds to avoid the soot formation as evidenced by the Raman spectra (see the supplementary material, Sec. S3).
Following the irradiation, final activation annealing (post-annealing) is done in a tube furnace in nitrogen at a pressure of 80 mbar without a flow, at a temperature of 850 \textdegree C for 30 minutes. 
The samples are placed in a graphite box (GB, unless stated otherwise) inside the furnace to ensure the presence of carbon in the environment. 
This is the so-called standard process\cite{Fischer2021ControlledEngineeringb} we use for emitter generation as schematically represented in Figure~\ref{fig:fig1}(b). Indicated batches undergo post-annealing in a quartz boat (QB).

Some batches of samples undergo an additional annealing step after the exfoliation, prior to any oxygen irradiation or post-annealing steps (see Figure~\ref{fig:fig1}(a)). This so-called \lq pre-annealing\rq ~process is done in an argon environment with 10$\%$ (v/v) oxygen at a flow rate of 300 sccm, while maintaining a pressure of approximately 0.7 atm, at a temperature of 700\textdegree C. The temperature ramp rate for heating and cooling is set at 40°C/min, with the entire process spanning roughly two hours. This annealing step is meant to remove organic contaminants on or around the hBN flakes\cite{Neumann2023OrganicMica}. 

In sections \ref{subsec:OxygenIrradiation}, \ref{subsec:CarbonAnneal}, and \ref{subsec:OxygenPre-anneal}, we show results for bulk hBN crystals purchased from the commercial supplier HQGraphene, referred to as \lq Type I\rq samples. We also studied hBN flakes exfoliated from crystals synthesized in a high-pressure, high-temperature process (HPHT) from the National Institute for Materials Science (NIMS)\cite{Watanabe2004Direct-bandgapCrystal, Taniguchi2007SynthesisSolvent} in experiments involving oxygen pre-annealing, referred to as \lq Type II\rq ~samples in this work.
In section \ref{subsec:Comparsion}, we distinguish the generated emitter densities between these two types of hBN for various processing methods involving oxygen pre-annealing. The batch names, the treatments undergone by each batch, and the corresponding emitter densities are also listed in Table~\ref{tab:table1}. 

\begin{figure*}
    \centering
    \includegraphics[width=0.9\linewidth]{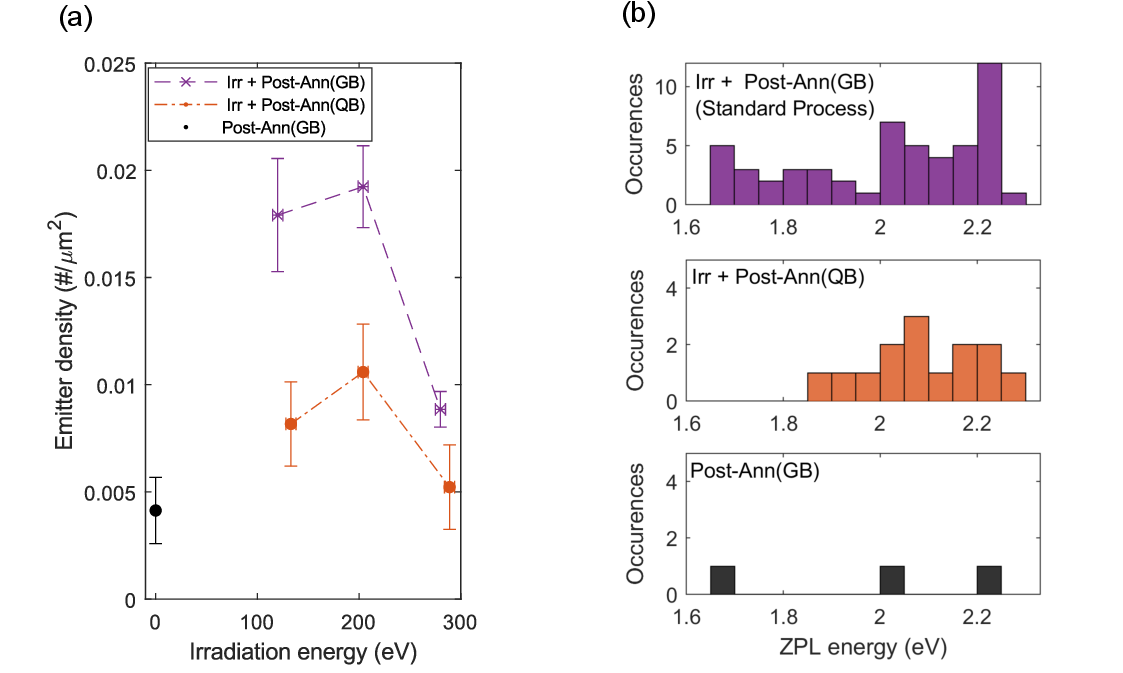}
    \caption{(a) Density of emitters plotted against the irradiation energy for the three different processes as indicated. Irr + Post-Ann(GB), shown in purple markers, indicates the standard process in which the samples undergo irradiation at indicated energies at room temperature, followed by nitrogen annealing in a graphite box (A1). Irr + Post-Ann(QB) shown in orange markers indicates the irradiated samples undergoing nitrogen annealing in a quartz boat instead of a graphite box (A2). Post-Ann(GB), shown in black, stands for a sample without irradiation undergoing graphite box annealing in nitrogen directly after exfoliation (A3). In all cases, the annealing is done at a fixed pressure of around 80 mbar without a flow. (b) Distribution of ZPL energy of all the observed emitters in the indicated processes.}
    \label{fig:fig2}
\end{figure*}

\subsection{Optical characterization}
For the photoluminescence measurements, a laser at 516 nm excites the samples, and a 50X microscope objective (0.6 NA) collects the emitted light from the samples filtered by a long-pass filter at 532 nm to the spectrometer or to the single-photon detectors during the intensity correlation measurements. All measurements are performed at room temperature. The calculation of the areal number density of emitters is discussed in detail in the supplementary material, Sec. S4.

Figure~\ref{fig:fig1}(c-e) shows the optical image and PL map with a grid of 0.5$\mu$m of representative flakes. The PL  map shows the normalized integrated photoluminescence counts from 1.5 to 2.29 eV. Figure~\ref{fig:fig1}(f-h) shows spectra for several quantum emitters. 
We identify a quantum emitter by having a relatively sharp peak in the spectrum in the range between 1.6 and 2.3 eV. For the second-order autocorrelation measurement g\textsuperscript{(2)}($\tau$), the spectral region containing the ZPL (grey area) is sent to the Hanbury Brown-Twiss measurement setup. We find clear antibunching around $\tau=0$.
From the fits\cite{Fischer2021ControlledEngineeringb} we derive the values of g\textsuperscript{(2)}(0), as indicated in the figures~\ref{fig:fig1}(i-k).
In supplementary material, Sec. S1, we show representative emitter spectra in different batches generated by the various treatments discussed in this work. 
The time traces of selected representative emitters are shown in the supplementary material, Sec. S2.

\begin{figure*}
    \centering
    \includegraphics[width=0.9\linewidth]{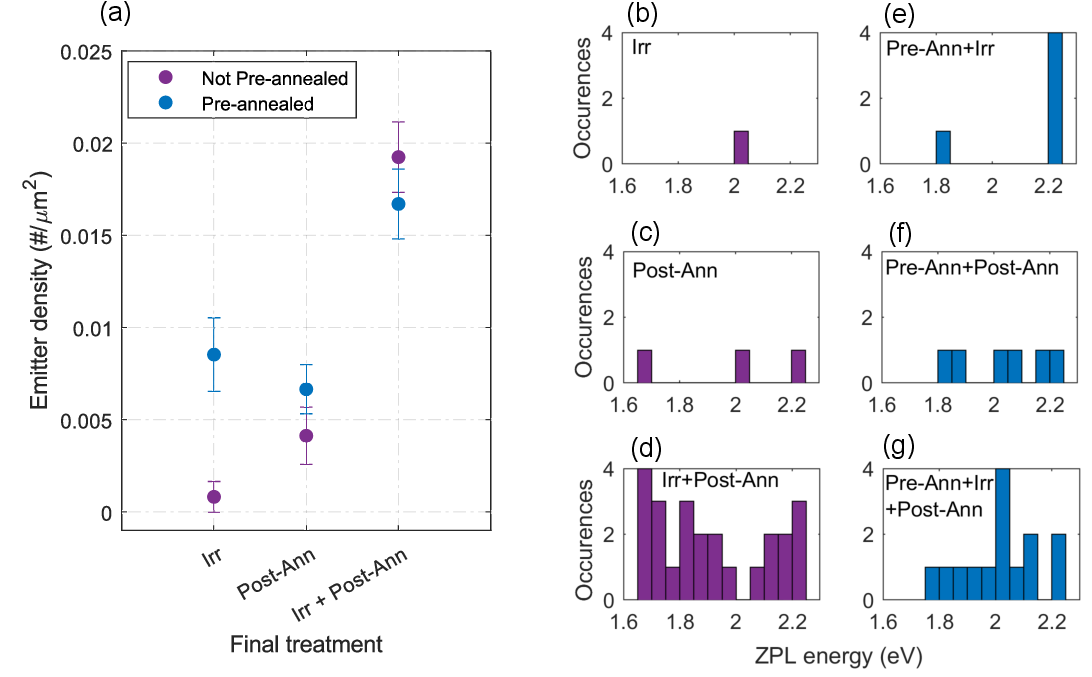}
    \caption{(a) Comparison of emitter densities in samples with and without pre-annealing in Type I hBN samples. Blue markers indicate the samples that have undergone pre-annealing treatment, and the purple ones indicate those that have not. The x-axis represents the final treatment undergone by the sample. (b) - (g) shows the ZPL energy distribution for the samples indicated in each plot. Note here that all indicated irradiation is at $\sim$ 204 eV, and all post-annealing is in a graphite box.}
    \label{fig:fig3}
\end{figure*}

\section{Results}\label{sec:Results}

\subsection{Oxygen irradiation}\label{subsec:OxygenIrradiation}
Figure \ref{fig:fig2}(a) shows the clear influence of the oxygen irradiation in the generation of emitters in hBN. All samples in the figure are prepared in the same way (see Methods) but with varying kinetic energies of oxygen irradiation.
In the case of no oxygen irradiation, we observe an emitter density of 0.004$\pm$0.002 $\#/\mu m^2$.
This value is significantly lower than the emitter density for the cases with irradiation at kinetic energies of 120 eV, 204 eV, and 280 eV, which yield 0.018$\pm$0.003, 0.019$\pm$0.002, and 0.009$\pm$0.001 $\#/\mu m^2$, respectively.
The highest yield is obtained with 204 eV and shows that a combination of oxygen irradiation and post-annealing generates a larger amount of emitters than a simple activation with annealing only. This is in line with our previous study~\cite{Fischer2021ControlledEngineeringb}, which discussed the fact that the oxygen irradiation amorphizes the top few layers and the subsequent inert annealing recrystallizes the layers, leading to the possible formation of optically active defects.

\subsection{Carbon-rich annealing environment}\label{subsec:CarbonAnneal}
Figure \ref{fig:fig2}(a) also displays the comparison between two different environments for the post-annealing step.
The purple curve, showing the emitter density for post-annealing in the carbon-rich graphite box, indicates a significantly larger emitter density than the carbon-poor quartz boat (orange curve). This difference of approximately a factor of 2 reveals a clear significance of the presence of carbon in the annealing environment.

Figure \ref{fig:fig2}(b) shows that the distribution of ZPL energies is much broader for post-annealing in a carbon-rich environment, indicating emitters with different microscopic structures. 
This could have several causes. A large number of carbon impurities in the hBN lattice have been proposed, (V\textsubscript{B}C\textsubscript{N}\textsuperscript{-}, C\textsubscript{2}C\textsubscript{B}, C\textsubscript{2}C\textsubscript{N}, C\textsubscript{B}C\textsubscript{N}C\textsubscript{B}C\textsubscript{N}\cite{Mendelson2021IdentifyingNitride, Fischer2023CombiningOrigin, Kumar2023LocalizedNitride}), which all have different ZPL energies. Alternate explanations could include the creation of graphene quantum dots embedded in the hBN lattice in a carbon-rich environment\cite{Li2011GrapheneSheets, Chen2019Sub-10Nitride}, or the presence of PAH molecules\cite{Neumann2023OrganicMica} pinned to the defect sites on the hBN\cite{deHaas2024ChargeNitride, Han2021PhotostableTemperatureb, Smit2023SharpSurface}.

\subsection{Oxygen pre-annealing}\label{subsec:OxygenPre-anneal}

In this section, we discuss the influence of pre-annealing in a molecular oxygen environment applied prior to irradiation and post-annealing processes. This pre-annealing step aims to remove any organic residues on and around the hBN surface that might create emitters. This approach is inspired by the work of Neumann et al. \cite{Neumann2023OrganicMica}, which reported completely dark emission in such flakes even after inert post-annealing, suggesting a correlation between organic residues coming from the micromechanical exfoliation of hBN crystals and emitter formation. We observed a low emitter density of 0.009$\pm$0.002 $\#/\mu m^2$ in flakes undergoing oxygen pre-annealing alone, with some exhibiting single-photon behavior (Figure~\ref{fig:fig1}(f) and (i)). Figure \ref{fig:fig3}(a), (b-g) illustrate the influence of oxygen pre-annealing, comparing emitter densities and ZPL energy distribution for different treatment combinations. Our results show that the standard process (Irr + Post-Ann) generates a similarly high emitter density of 0.017$\pm$0.002 $\#/\mu m^2$, regardless of whether the oxygen pre-annealing step is performed or not. This indicates that oxygen pre-annealing does not significantly influence emitter density, and that irradiation is the dominant process for emitter generation in hBN.

We find some differences in our methods and measurement results compared to those reported by Neumann et al. \cite{Neumann2023OrganicMica}, the most notable being the observation of a non-zero emitter density in pre-annealed samples. There is a possibility that high-temperature annealing in an oxygen atmosphere activates some lattice defects, as shown in previous works \cite{Chen2021GenerationNitride, Mohajerani2024NarrowbandNitride, Liu2023SingleBeam}. This observation could also be caused by residual organic molecules from solvents or tape residue that cannot be removed by oxygen annealing alone. Molecules beneath the flake are shielded from oxygen gas, and the heat may cause them to form fluorescent PAHs \cite{Neumann2023OrganicMica}. However, we observe a significant increase in emitter density after irradiation, suggesting that crystallographic defects could play an important role in emitter formation. Some emitters found after the standard process may result from organic contamination during fabrication steps, which may not be entirely removed even after undertaking an oxygen pre-annealing step. Nonetheless, the consistently higher emitter density observed with our standard process suggests an important role of defects in the generation of emitters in hBN. It remains unclear whether these defects act as emitters themselves, serve as pinning sites for molecules, or function as growth sites for graphene quantum dots within the lattice. In the future, incorporating additional cleaning with solvents that can eliminate tape residues without leaving their own residues, similar to the method used by Neumann et al. \cite{Neumann2023OrganicMica}, could help to rule out the possibility of molecules beneath or pinned on surface defects induced by irradiation.

\subsection{Comparison between different hBN types}\label{subsec:Comparsion}
\begin{figure}
    \centering
    \includegraphics[width=0.9\linewidth]{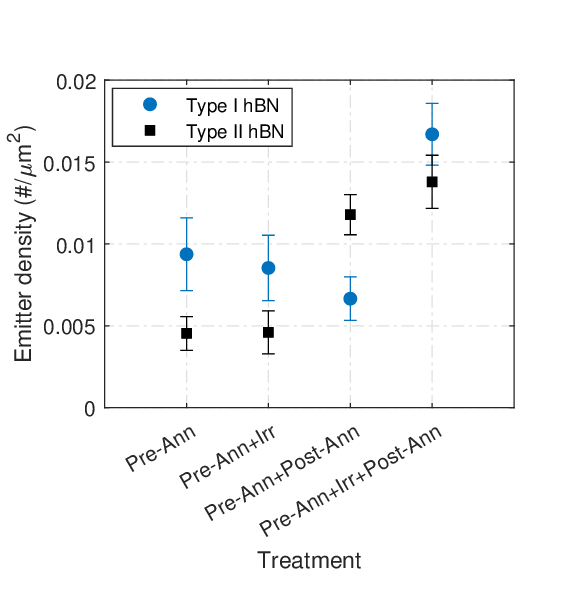}
    \caption{The density evolution of emitters on pre-annealing only (B1), pre-annealing followed by irradiation (B2), pre-annealing followed by post-annealing in nitrogen in graphite box (B3), pre-annealing followed by irradiation and post-annealing (B4). The blue circles are the results from Type I hBN flakes, and the black squares are those from Type II hBN flakes.}
    \label{fig:fig4}
\end{figure}

Last, we compared hBN from HQGraphene (referred to as Type I hBN) and crystals synthesized in a high-pressure, high-temperature process\cite{Watanabe2004Direct-bandgapCrystal, Taniguchi2007SynthesisSolvent} (referred to as Type II hBN) as shown in Figure~\ref{fig:fig4}.
Type II hBN is of higher crystallographic quality than Type I hBN and should contain less carbon contamination \cite{Taniguchi2007SynthesisSolvent, Schue2017CharacterizationLayers, Su2024FundamentalsTutorial}.
Without applying the standard process of generating quantum emitters, we find that Type I hBN has twice as large an emitter density as Type II hBN. However, when the standard process of irradiation and post-annealing is applied, we find similar emitter densities of 0.014$\pm$0.002 \#/$\mu m^2$, with that of Type I hBN being slightly higher.

The low impurity concentration of Type II hBN aligns well with the standard scenario of defect formation discussed earlier. This observation can be linked to the contribution of optically active defect centers. At the same time, it may also suggest the possibility of a molecular origin supported by the presence of defects. Low-quality hBN flakes, due to their low crystallinity, may have more cracks or edges where molecular species can integrate. There is a higher chance that the organic residues from the exfoliation tape or from any fabrication step remain in these regions. Consequently, the likelihood of molecular formation would also be higher in these low-crystalline flakes, as even oxygen pre-annealing may not be sufficient to remove all possible trapped tape or organic residues in these edges or cracks.

\section{Conclusion}\label{sec:Discussion}
In summary, we have shown the influence of different fabrication methods on the formation of quantum emitters in hBN. 
We found that (i) oxygen irradiation has a strong influence on the emitter density, with an optimal kinetic energy of 204 eV. (ii) Post-annealing in the graphite box generates an increased emitter density by approximately a factor of 2 as compared to that in the quartz boat, with a wider distribution of zero-phonon line toward the (infra)-red. (iii) Oxygen pre-annealing does not have a strong influence on the emitter density, with the density being almost the same for the flakes treated by the standard process. (iv) hBN from HQGraphene shows a slightly higher emitter density compared to the more pristine NIMS hBN.

The influence of oxygen irradiation and the increased density from annealing in a carbon-rich environment would suggest carbon-based defects. 
Crystallographic defects do not have to be the emitters themselves, as they may act as pinning sites that attract organic molecules from the environment\cite{Neumann2023OrganicMica}. 
The presence of polycyclic aromatic hydrocarbons or (nano)graphene on top of the flake can therefore not be excluded. 
Besides, there may still be organic residue underneath the flake that can not be
removed in the oxygen annealing treatment. 
However, the higher emitter density in irradiated batches compared to non-irradiated batches suggests that the crystallographic defects play a significant role, whether as quantum emitters\cite{Tran2016QuantumMonolayers, Mendelson2021IdentifyingNitride, Fischer2023CombiningOrigin, Kumar2023LocalizedNitride}, growth sites for embedded graphene quantum dots\cite{Li2011GrapheneSheets, Chen2019Sub-10Nitride}, or pinning sites for PAHs\cite{Smit2023SharpSurface, Han2021PhotostableTemperatureb, deHaas2024ChargeNitride}. 

Reducing the number of unintended emitters originating from naturally occurring defects or organic contaminants enhances our understanding of the irradiation process. This could be achieved by using more pristine hBN, as shown in Figure \ref{fig:fig4}. 
Additionally, we could apply a solvent that can clean beneath the flake, where oxygen annealing has no effect, while leaving no residue of its own. 
Improving the ratio between intentional and unintentional emitters is crucial for developing effective generation methods and controlled sample fabrication for quantum technologies.

\begin{acknowledgments}

We thank Tim Booth, Michael Neumann, and Iris Niehues for helpful discussions. P.C.S., M.W., and N.S. acknowledge the Independent Research Fund Denmark Natural Sciences (Grant No. 0135-00403B).
M.W., N.S., and S.X. acknowledge the support by the Danish National Research Foundation through the NanoPhoton Center for Nanophotonics, Grant No. DNRF147 and the Center for Nanostructured
Graphene, Grant No. DNRF103.  P.C.S., C.K., and N.S. acknowledge the
Novo Nordisk Foundation NERD Programme, project QuDec
NNF23OC0082957. S.X. acknowledges the support by the Villum Foundation (Project No. 70202) and the Independent Research Fund Denmark (Grant No. 2032-00351B).
K.W. and T.T. acknowledge support from the JSPS KAKENHI (Grant Numbers 21H05233 and 23H02052), the CREST (JPMJCR24A5), JST and World Premier International Research Center Initiative (WPI), MEXT, Japan.
J.M.C. and J.A.D.-N. thank the support by Junta de Castilla y Leon  under the Research Grant number SA103P23, the Ministry of Science and Innovation under the grants PID2021-128154NAI00, RYC2019-028443-I, RYC2023-044965-I and CNS2024-154588, as well as the USAL-NANOLAB for the use of clean room facilities. J.A.D.-N. thanks the support from the Universidad de Salamanca for the Maria Zambrano postdoctoral grant. J.M.C. also acknowledges financial support of the European Research Council (ERC) under Starting grant ID 101039754, CHIROTRONICS, funded by the European
Union. Views and opinions expressed are, however, those of the author(s) only and do not necessarily reflect those of the
European Union or the European Research Council. Neither the European Union nor the granting authority can be held responsible for them. 

\end{acknowledgments}

\section*{AUTHOR DECLARATIONS}
\subsection*{Conflict of Interest}
The authors have no conflicts to disclose.

\section*{Author Contributions}
N.S., P.C.S., J.M.C., M.F., M.W., and S.X. conceived the project. P.C.S., J.M.C., J.A. D.-N., prepared the samples by irradiation with oxygen and pre-annealing. K.W. and T.T. synthesized the hBN crystals. P.C.S. and C.K. optically characterized the prepared samples with photoluminescence, second-order correlation, and time-resolved measurements. D.I.M. and S.C. helped in the post-annealing sequence of the preparation of the samples. P.C.S. and C.K. analyzed the results. All authors contributed in the discussion of the results and the writing of the article.

\section*{DATA AVAILABILITY}
The data that support the findings of this study are available from the corresponding author upon reasonable request.

\section*{References}
\bibliography{references}

\onecolumngrid

\section*{Supplementary information for:\\Systematic investigation of the generation of luminescent emitters in hBN via irradiation engineering}

\renewcommand{\figurename}{Fig.}
\renewcommand\thefigure{S\arabic{figure}} 
\renewcommand{\tablename}{TABLE.}
\renewcommand\thetable{S\arabic{table}}

\renewcommand\thesection{S\arabic{section}} 
\setcounter{section}{0}
\setcounter{figure}{0}
\setcounter{table}{0}

\section{More representative emitter spectra}\label{sec:Appendix_MoreSpectra}
\begin{figure*}[h]
    \centering
    \includegraphics[width=0.6\linewidth]{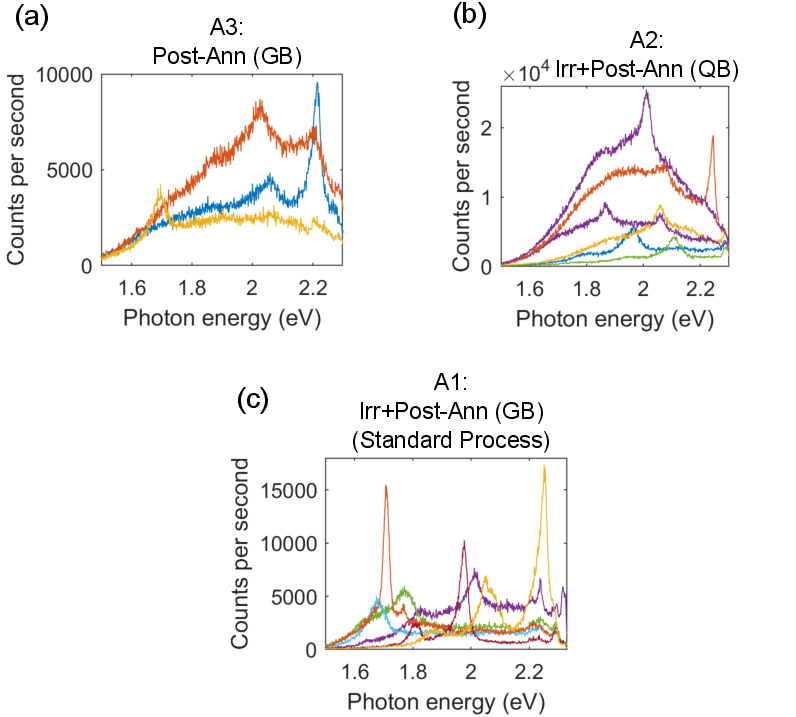}
    \caption{Representative emitter spectra of emitters generated in (a) Batch A3, the
one in which flakes are directly post-annealed in graphite box (Post-Ann(GB)) (b) Batch
A2, which is irradiated followed by post-annealed in quartz boat (Irr+Post-Ann(QB)) (c)
Batch A1 undergoing the standard process, which is irradiation followed by post-annealing
in graphite box (Irr+Post-Ann(GB)). All mentioned irradiation is done at $\sim$204 eV.}
    \label{fig:SI_EmittersBatchA}
\end{figure*}

\begin{figure*}
    \centering
    \includegraphics[width=0.6\linewidth]{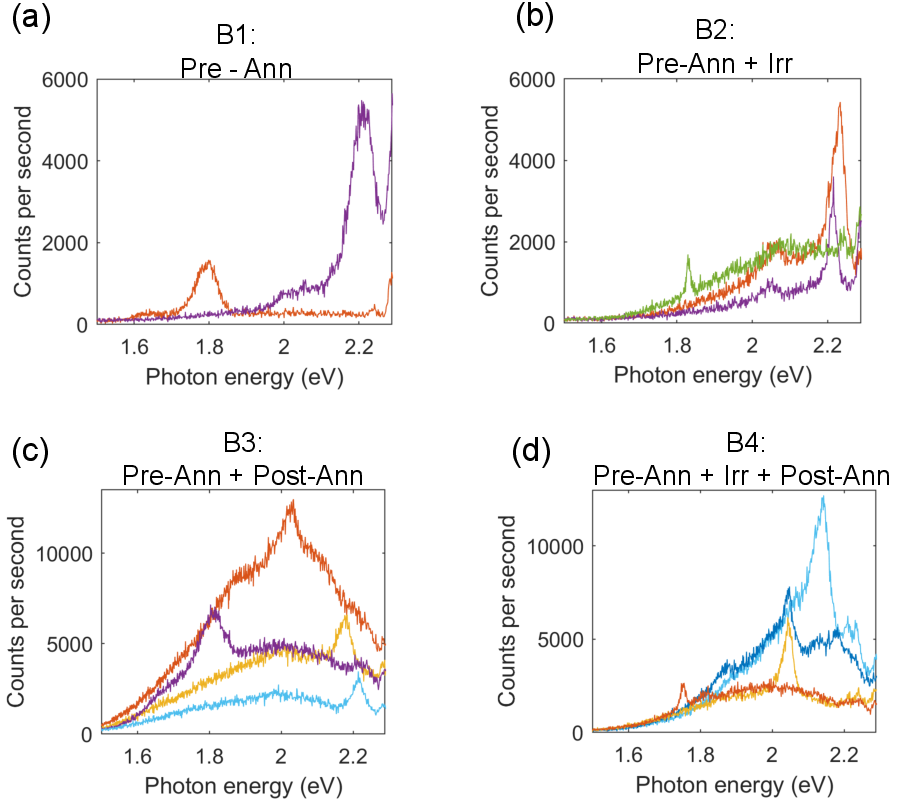}
    \caption{Representative spectra of emitters generated in batches undergoing the pre-annealing process (a) Batch B1 with only the pre-annealing process. (b) Batch B2 undergoing
subsequent irradiation. (c) Batch B3 undergoing the post-annealing treatment without an
irradiation. (d) Batch B4 undergoing the standard process of irradiation followed by post-annealing. All mentioned irradiation is done at $\sim$204 eV, and all mentioned post-annealing is done in the graphite box.}
    \label{fig:SI_EmittersBatchB}
\end{figure*}

\clearpage

\section{Photophysics and stability of selected emitters}\label{sec:Appendix_Photophysics}
\begin{figure*}[h]
    \centering
    \includegraphics[width=0.9\linewidth]{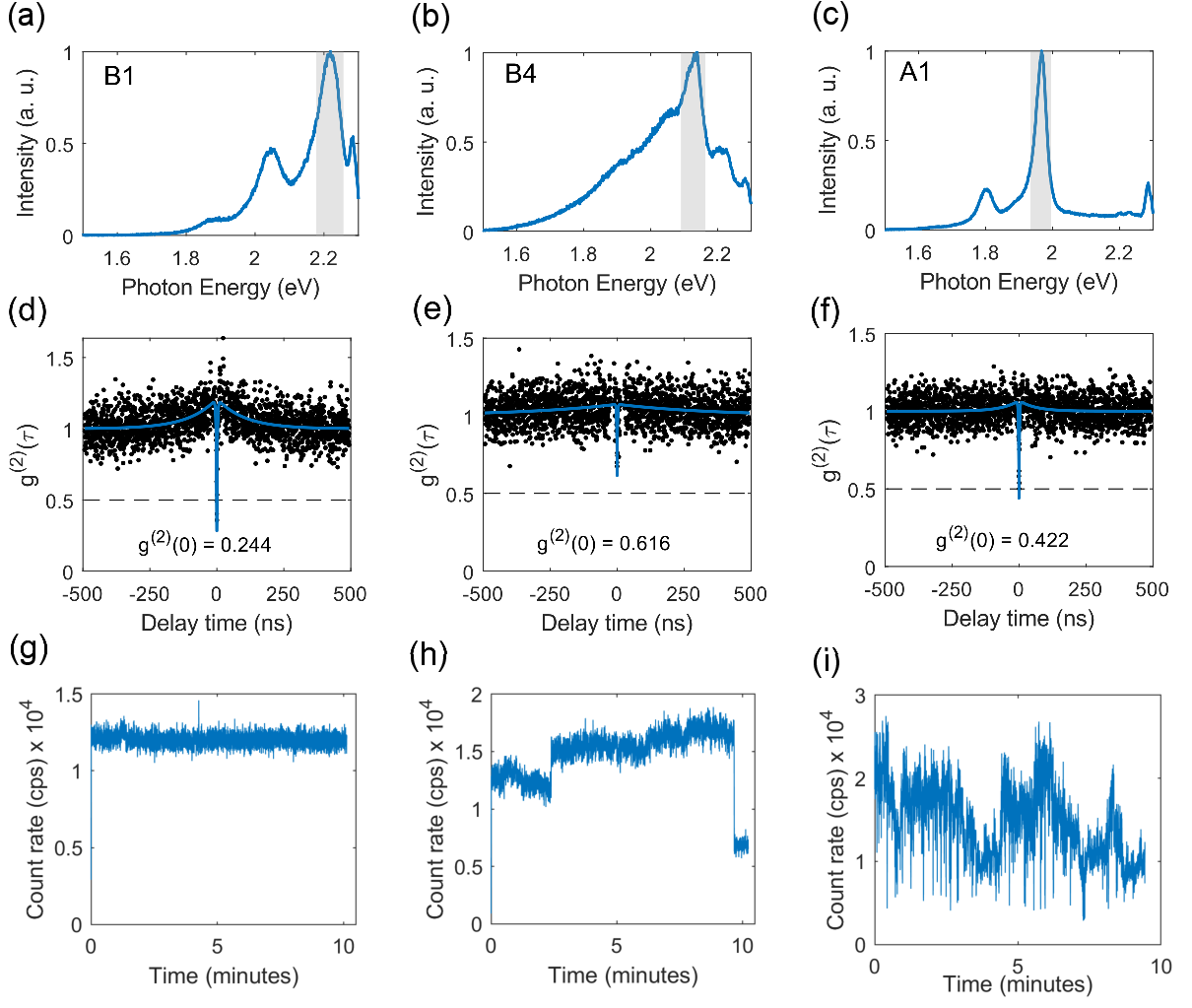}
    \caption{Photoluminescence spectrum of a representative emitter generated by (a) pre-annealing alone (B1), (b) pre-annealed followed by standard process (B4), (c) the standard process (A1). (d-f) g\textsuperscript{(2)} function of the emitters shown in (a-c) from the spectral region shaded in grey. (g-i) Time trace of the corresponding emitters shown in (a-c) on continuous excitation over a period of 10 minutes.}
    \label{fig:fig5}
\end{figure*}

\clearpage
\section{Raman signals from hBN undergoing irradiation and post-anneal treatment}\label{sec:Appendix_Raman}
\begin{figure*}[h]
    \centering
    \includegraphics[width=\linewidth]{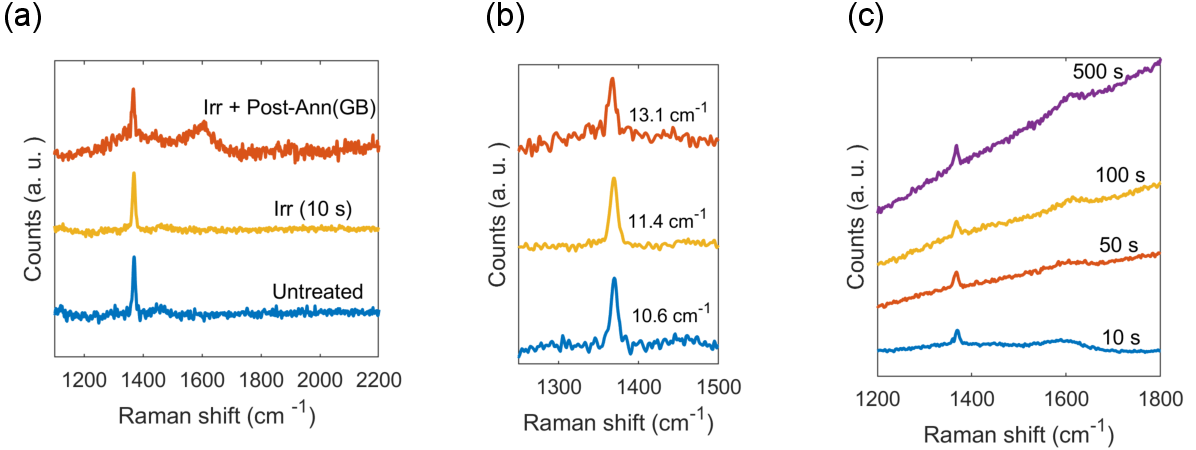}
    \caption{(a) Raman signal from an exfoliated hBN flake (untreated), oxygen irradiated for 10 seconds (Irr (10s)), and subsequently annealed in graphite box (Irr+Post-Ann(GB)). The band around 1600 cm\textsuperscript{-1} arises from carbon in the sample\cite{Hou2018LocalizedNitride}, possibly during the annealing in the carbon environment. (b) The zoom-in of E\textsubscript{2g} Raman mode of hBN with the FWHM value of each peak, obtained from fitting with a Gaussian. The E\textsubscript{2g} peak value for untreated, irradiated, and irradiated + post-annealed flakes is obtained at 1369.7, 1369.2, 1366.7, respectively. The increase in FWHM value is an indicator of damage in the lattice during each stage, or intuitively the formation of defects as discussed in Ref. \citenum{Schue2017CharacterizationLayers}. (c) Damage caused by irradiation on hBN flakes with time as evidenced from the Raman signals. The signals are measured from samples undergoing irradiation followed by post-annealing in a graphite box, with an irradiation duration as indicated for each spectrum. A longer duration of irradiation causes a huge background luminescence. The blue spectra in (c) correspond to the orange in (a) and (b).}
    \label{fig:SI_Raman}
\end{figure*}

\section{Area number density calculation}\label{sec:Appendix_EmitterDensity}

For all the emitter statistics presented, we used emitter density per area of the flakes, similar to the calculation in Ref \citenum{Fischer2021ControlledEngineeringb}. The geometrical area of the flakes is calculated from the sample's optical image. An error of 5 $\mu$m\textsuperscript{2} is assigned for determining the flake area.
The areal number density, 
\begin{equation}
    D = \frac{N}{A}, 
\end{equation}
where N  is the total number of emitters in \lq n\rq number of flakes and \lq A\rq is the total area of \lq n\rq number of flakes.

The error in determining the density is given by,

\begin{equation}
  \Delta D = \frac{\Delta N}{A} + \frac{N.\Delta A}{A^2},
\end{equation}
where 
\begin{equation}
    \Delta A = n.5 \quad\mu m^2
\end{equation}
and
\begin{equation}
  \Delta N =\begin{cases}
    1, & \text{if $N<4$}.\\
    \text{round(N/10)}, & \text{if $N\geq5$}.
  \end{cases}
\end{equation}
    The function round(x) rounds off x to the nearest integer. 

The following tables show all the details of the flake area and emitters, which are the basis of our statistical study.

\begin{table}
  \caption{Data for emitter density calculation in Type I samples undergoing the standard process (Batch A1), for the indicated irradiation powers, and the non-irradiated sample undergoing direct post-annealing in graphite box (Batch A3) discussed in Section 
   III A of the Main text.}
  \label{tbl:Table_Appendix1}
  \begin{tabular}{lllll}
    \hline
    Irradiation Power (W)  & Irradiation energy (eV) & Flake\ area($\mu m^2$) & Flakes & Emitters  \\
    \hline
    20& 120 $\pm$5 & 893.36 &4& 16 \\
    40& 204$\pm$5  & 1247.84 &4& 24 \\
    60& 280$\pm$5  & 1469 &5& 13 \\
    Non-irradiated (Batch A3)& 0 &726.36 & 6&3 \\

    \hline
  \end{tabular}
\end{table}


\begin{table}
  \caption{Data for emitter density calculation in Type I samples undergoing the irradiation followed by post-annealing in quartz boat (Batch A2) discussed in Section 
  III B of the Main text, for the indicated irradiation powers.}
  \label{tbl:Table_Appendix2}
  \begin{tabular}{lllll}
    \hline
    Irradiation Power (W)  & Irradiation energy (eV) &Flake\ area($\mu m^2$) & Flakes & Emitters  \\
    \hline
     20& 133 $\pm$5& 612.21 &5& 5 \\
     40& 204$\pm$5  &566.63 &5& 6 \\
     60& 289$\pm$5 &574.7 &5& 3 \\
    \hline
  \end{tabular}
\end{table}

\begin{table}
  \caption{Data for emitter density calculation in Type I samples undergoing the indicated treatments as discussed in Sections 
  III C and III D of the Main text.}
  \label{tbl:Table_Appendix3}
  \begin{tabular}{lllll}
    \hline
   Batch & Treatment  & Flake\ area($\mu m^2$) & Flakes & Emitters  \\
    \hline
    B1 & Pre-Ann only & 533.69 &4& 5\\
     B2 &Pre-Ann + Irr & 586.1 & 4&5\\
     B3 &Pre-Ann + Post-Ann & 901.53 & 6&6\\
     B4 &Pre-Ann + Irr + Post-Ann &838.62 & 7&14\\
    \hline
  \end{tabular}
\end{table}


\begin{table}
  \caption{Data for emitter density calculation in Type II hBN samples undergoing the indicated treatments as discussed in 
  Section III D of the Main text.}
  \label{tbl:Table_Appendix4}
  \begin{tabular}{lllll}
    \hline
    Batch & Treatment  & Flake\ area($\mu m^2$) & Flakes & Emitters  \\
    \hline
    B1 &Pre-Ann & 1103.05 &6& 5 \\
    B2 &Pre-Ann + Irr & 870.23 & 6&4\\
    B3 &Pre-Ann + Post-Ann & 1103.05 & 6&13\\
    B4 &Pre-Ann + Irr + Post-Ann &870.23 & 6&12\\
    \hline
  \end{tabular}
\end{table}

\end{document}